# Magnetic relaxation of a system of superparamagnetic particles weakly coupled by dipole-dipole interactions

Pierre-Michel Déjardin

*Laboratoire de Mathématiques et Physique, Université de Perpignan Via Domitia, 52, Avenue Paul Alduy, 66860 Perpignan Cedex, France*

**ABSTRACT**

The effect of long range dipole-dipole interactions on the thermal fluctuations of the magnetization of an assembly of single-domain ferromagnetic particles is considered. If orientational correlations between the particles are neglected, the evolution of the magnetization orientations may be described by a nonlinear Fokker-Planck equation (FPE) reducing to the usual linear one in the limit of infinite dilution [W.F. Brown Jr, Phys. Rev. **130**, 1677 (1963)]. The thermally activated relaxation time scale of the assembly is estimated, leading to a simple modification of the axially symmetric asymptotes for the superparamagnetic relaxation time.

# I. INTRODUCTION

Fine magnetic particles are characterized by thermal instability of the magnetization [1] resulting in superparamagnetism [2]. The thermal fluctuations and relaxation of the magnetization **M**(*t*) are relevant in information storage, biomedical applications and rock magnetism etc. The treatment of the thermal fluctuations in superparamagnets begun by Néel[1] was further developed by Brown [3, 4] He treated them using the theory of the classical Brownian motion using the Fokker-Planck equation (FPE) for the probability density function *W* of magnetization orientations on the unit sphere. Thus he derived approximate expressions for the superparamagnetic relaxation time both in the low and high energy barrier approximations. Moreover the various responses to externally applied fields may be calculated [5]. However, this treatment ignores concentration effects on the relaxation time because inter-particle interactions are neglected [6]. Now magnetic relaxation as modelled by Brown is, in many respects, a replica of the Debye theory of dielectric relaxation in dipolar liquids. Hence, the analogy is useful in modelling the effect of dipole-dipole interactions on superparamagnetic relaxation.

Dipolar interactions in the orientational dynamics of an assembly of dipoles always lead to involved calculations as is apparent in dielectric relaxation of polar fluids [7,8], where much modelling has already been done. The first Debye-Lorentz calculation of the static permittivity valid for polar gases at very low densities was improved upon by Onsager [8], which, as outlined by van Vleck [9], can be transposed to assemblies of Langevin paramagnets, Onsager in order to improve Debye's formula accounted for the reaction of a given polar molecule on its surroundings by supposing that a typical dipole of an assembly resided at the center of a macroscopic spherical cavity in a dielectric medium, so generalizing the calculation of the static dielectric constant to higher densities. However, Onsager's model was criticized by Kirkwood [10] because it ignores the interactions between the dipole of the macroscopic cavity with its nearest neighbours. Consequently, Kirkwood, and later Fröhlich, have corrected Onsager's formula by introducing the "Kirkwood correlation factor", which accounts for the contribution of the neighbours of a molecule to the macroscopic dielectric permittivity [7,8].

However, the calculation of the *dynamic* permittivity of a dilute assembly of dipoles is far more involved than its static counterpart because, in general, the equation of motion of a typical dipole depends on the macroscopic properties that one is trying to calculate [7]. In effect, this task leads to extremely difficult calculations because the time dependence of the field involved in the equation of motion of a typical dipole is unknown. From an analytical point of view, various attempts to generalize Onsager's approach have been given which lead to permittivity values (e.g., the Onsager-Cole formula) that are incompatible with any relaxation time distribution as the corresponding Cole-Cole plots may lie outside of the Debye semicircle [7]. One approach to the problem is that of Nee and Zwanzig [11]. By assuming a delayed response between the dipole moment and the Onsager reaction field, hence allowing for a special type of memory effect (dielectric friction), they were able to rederive the Fatuzzo-Mason equation [7] for the dielectric permittivity $\varepsilon(\omega)$. A review of attempts to generalize the Kirkwood-Fröhlich theory to the dynamical case was given by Madden and Kivelson [12]. Furthermore, a more general extension to intermediate to large wavelengths was given by Bagchi and Chandra [13] who adapted the equations of generalized molecular hydrodynamics to dielectric relaxation.

Later, Zhou and Bagchi [14] have reconsidered Zwanzig's model [15] of dielectric relaxation of dipoles located at the sites of a cubic lattice. In particular, these authors pointed out the inadequacy of both the perturbation and the Nee-Zwanzig approaches to the collective dynamics, showing that these models for dielectric friction predict a too rapid decay of the dipole-dipole correlation function. Nevertheless, they still find Debye-like behaviour for the permittivity for relatively dilute dipolar systems, in agreement with both the perturbation and continuum approaches.

The time dependence of the field involved in the equations of motion being unknown, Coffey [16] considered the reaction field to act on a typical dipole as an external field of large amplitude, and calculated the polarisation decay using Picard's method of successive approximations for solving integral equations. Like for Zwanzig's model [15], this calculation demonstrates that the complex polarizability is governed by an infinite *discrete* set of microscopic relaxation mechanisms. The advantage in Coffey's method is that the many-body problem is bypassed entirely, and the distribution of relaxation times arise naturally by including a potential in the Smoluchowski equation. However, the extension of this calculation to superparamagnetic relaxation is by no means a trivial matter.

Yet another model that has been introduced in the dielectric relaxation literature is Berne's forced rotational diffusion model [17]. Inspired by the literature on dielectric relaxation in electrolytes [18], Berne treated the rotational Brownian motion of interacting electric dipoles self-consistently so that the FPE describing the orientational dynamics of the polar molecules is *nonlinear*. The difficulty in handling this FPE is mainly a mathematical one, which Berne overcame by linearizing it about statistical equilibrium *before* taking its spatial Fourier transform, in the spirit of the Debye-Hückel theory. Then he took the zero wave-vector limit of the resulting equation in order to obtain the relevant correlation functions for the calculation of the dielectric constant of the dipolar assembly. This model in its nonlinear version has been also considered by Warchol and Vaughan [19].

However, analytical methods for modelling dipole-dipole interactions in the *dynamical* properties of interacting single-domain ferromagnetic particles have been much less well developed than their dielectric relaxation counterparts. The first attempt to estimate the relaxation time of the magnetization in the presence of dipolar interactions is due to Shtrikman and Wohlfarth (SW) [20]. They predicted that inter-particle interactions would *increase* the magnetic stability of the system against thermal agitation. This result motivated a number of experimental and theoretical studies on magnetic relaxation in fine particle systems by altering the amount of magnetic matter in the studied samples [6]. Now the SW estimation relies on the calculation of the relaxation time of a *single* magnetic moment in the mean field of the others, and their method resembles subsequent approaches proposed in the literature [21,22,23,24]. However, these procedures do not yield the total magnetic moment as they are valid for non-interacting particles only. Instead, in order to calculate the dynamical features of the assembly, one must consider the dynamics of *the vector sum of all magnetic moments*.

From an experimental point of view, the properties of single-domain ferromagnetic particle assemblies have been classified according to the interaction strength (see, e.g., Ref. 25). To summarize, the behaviour of assemblies may be separated into four categories : (i) pure superparamagnetism, where the relaxation time has Néel-Brown behaviour below the blocking temperature ; (ii) modified superparamagnetism, where the interactions are still too weak to produce a collective state, however the Arrhenius law is still formally obeyed and the memory effects observed in the zero-field cooled magnetization (ZFC) experiments, typical of a collective state are absent ; (iii) the superspin glass phase where memory effects are present and where, clearly, orientational correlations can no longer be neglected ; here, the interactions are moderate, the particle size distribution should be rather narrow and the locations of particles in space must be random in order to produce frustration ; and (iv) the superferromagnetic state, where the interactions are so strong that magnetic order may appear. From the point of view of theoretical modelling, the foregoing classification is particularly useful.

Now, it is clear that the understanding of nonequilibrium properties of such assemblies commences with the modelling of their equilibrium properties. In particular, this has been accomplished by Jonsson and Garcia-Palacios [26] by adapting the Rosenberg-Lax calculation of equilibrium properties of electric dipoles at the sites of a crystalline lattice [27]. This method was extended by Kachkachi and Azeggagh [28] when an external dc field is applied to the assembly. In particular, the Kachkachi-Azeggagh calculation shows that the approach to saturation is still governed by a $1-b/H$ law, where $H$ is the applied field and $b$ is a constant, although the equilibrium magnetization curves appreciably deviate from Langevin superparamagnetic behaviour at intermediate fields. These calculations include interactions to second order.

In order to visualize the effect of dipole-dipole interactions on superparamagnetic relaxation, the dynamics of systems of two coupled spins subjected to thermal agitation has been studied. We mention here the work of Rodé *et al*. [29] who, by modelling the motion of two anisotropic spins using the two-body Smoluchowski equation for rotation in the plane, have obtained a simple relaxation time expression in terms of an effective volume. They concluded that the effective volume varies from the true volume for very weak interaction to twice the particle volume for strong interactions, thereby leading to an increase of the magnetic stability and coercive force as the interaction strength increases. A similar conclusion was obtained by Lyberatos and Chantrell [30] who studied a three-dimensional version of the problem considered in Ref. 29 using Langevin dynamics simulations. The main difference between the works presented in Refs. 29 and 30 lies in the detailed dependence of the prefactor on the interactions because gyroscopic effects are included in Ref. 30. These results are compatible with SW's conclusions and are useful in anticipating the effect of dipole-dipole interactions on the relaxation time of an assembly. Nevertheless, in reality, the dynamics of the whole system of coupled particles has to be considered, as such interactions are long ranged, and, unlike exchange interactions, no cutoff interparticle distance is allowed in principle [6].

This problem has been tackled by a number of authors using diverse numerical approaches, yielding results that partly agree, and are partly at variance. Here, we must note the Monte-Carlo approach of Andersson *et al.* [31] who have focused mainly on the frequency and concentration behaviour on the temperature maximum of the ac susceptibility. They have found that this peak is shifted to higher temperatures as the frequency or concentration is increased, in qualitative agreement with experiment. However, gyromagnetic effects are not explicitly included in these calculations, prompting Berkov and Gorn [32] to numerically solve the set of Langevin equations for a system of spherical single-domain particles with uniaxial anisotropy and coupled by dipole-dipole interactions. This calculation was made without any assumptions regarding the strength of the interactions, and the focus is on the temperature and frequency behaviour of the imaginary part of the ac susceptibility of an assembly of nanoparticles. They found a crossover-type behaviour between assemblies with low anisotropy and high anisotropy, with a non-monotonous temperature behaviour of the susceptibility peak as a function of temperature and concentration at moderate to low damping. However, in most of these numerical studies, it is extremely difficult to ascertain the category of assembly to which these results apply. Moreover, in spite of the Berkov-Gorn objection to analyzing the experimental data in terms of energy barriers because it ignores the gyromagnetic effects, nevertheless the barrier-based approach is used to interpret experimental data [33]. Finally, we remark that even the second category of assemblies, (ii), mentioned in Ref. 25, i.e., superparamagnetic relaxation modified by interactions, has hardly ever been connected with Brown's original theory, i.e., starting with the Langevin equations of motion and accounting for the long range character of the dipolar torques. Moreover, the use of the known expression for the dipole-dipole interaction expression invariably assumes that the particles are always relatively far away from each other and also that they are all spherical [6].

The purpose of this paper is to propose a model for the magnetization relaxation of an assembly of single-domain ferromagnetic particles coupled via dipole-dipole interactions, by adapting Berne's model of forced rotational diffusion of electric dipoles to magnetic systems. This model allows straightforward reduction to Brown's model which is valid for infinitely diluted systems. Here we will suppose that orientational correlations may be neglected, so requiring one to solve a nonlinear Fokker-Planck equation which formally resembles that given by Brown. Thus, the calculations will be restricted to the case (ii), i.e., superparamagnetism and superparamagnetic relaxation as modified by dipole-dipole interactions.

## II. BASIC EQUATIONS

First, we present the steps that are necessary to compute the time-dependent magnetization of an assembly of single-domain ferromagnetic particles having a random spatial arrangement. We shall assume that the sample is infinite, that the particles all have the same volume and that their mechanical motion is frozen. We further assume that the interparticle interactions consist of dipole-dipole interactions only. The magnetic induction **B** in the sample arising from this interaction must satisfy Maxwell's equation

$$\mathrm{div}\,\mathbf{B}(\mathbf{r},t) = 0, \tag{1}$$

where

$$\mathbf{B}(\mathbf{r},t) = \mathbf{H}_{ddi}(\mathbf{r},t) + 4\pi\mathfrak{M}(\mathbf{r},t), \tag{2}$$

$\mathbf{H}_{ddi}(\mathbf{r},t)$ is the dipole-dipole magnetic field (the subscript "*ddi*" refers to dipole-dipole interaction) and $\mathfrak{M}(\mathbf{r},t)$ is the magnetic moment density. On defining the "volume magnetic charge density" by $\rho(\mathbf{r},t) = -\mathrm{div}\,\mathfrak{M}(\mathbf{r},t)$, we have the equation

$$\text{div}\,\mathbf{H}_{ddi}(\mathbf{r},t) = 4\pi\rho(\mathbf{r},t). \tag{3}$$

Furthermore, since we assume that no electric current flows through the sample, we have

$$\text{curl}\,\mathbf{H}_{ddi}(\mathbf{r},t) = \mathbf{0}, \tag{4}$$

and therefore

$$\mathbf{H}_{ddi}(\mathbf{r},t) = -\text{grad}\,\Psi(\mathbf{r},t) \tag{5}$$

where $\Psi$ is the magnetostatic scalar potential arising from the magnetic charge distribution. Therefore, Eqs. (3) and (4) are similar to the equations of electrostatics suggesting a simple transposition to magnetostatics. Clearly, Eqs. (3)-(5) may be combined in a single equation yielding

$$\Delta\Psi(\mathbf{r},t) = -4\pi\rho(\mathbf{r},t), \tag{6}$$

which is entirely analogous to the Poisson equation of electrostatics. Continuing this formal analogy, we may formally transpose Berne's reasoning in order to link the volume charge density with the microscopic concentration of particles as follows. Firstly, we have

$$\rho(\mathbf{r},t) = \sum_{i=1}^{N}\sum_{j\in i} Z_j \delta\left[\mathbf{r} - \mathbf{r}_i - s_j \mathbf{u}_i(t)\right], \tag{7}$$

where $N$ is the number of particles in the sample, $\mathbf{r}_i$ is the position of particle $i$, $\mathbf{u}_i(t)$ is the orientation of the magnetic moment of particle $i$, $\delta$ is the Dirac delta function and, by analogy with electrostatics, $Z_j$ is the magnetic charge (assumed located along the direction of the dipole moment of particle $i$ at some local abscissa $s_j$). In particular, we have $\sum_j Z_j = 0$ since there is no net global magnetic charge carried by particle $i$, and also we have $\sum_j s_j Z_j = M_s v_i$, the magnetic moment of particle $i$ with volume $v_i$. We further assume $v_i = v$ since we neglect the particle size distribution. Here, of course, $Z_j$ has only a formal physical meaning as magnetic charges do not exist. Instead, we are using magnetic charge only as a convenient mathematical tool, and as will be shown further, it will never occur explicitly in the calculations.

Following Berne again, we may introduce the linear magnetic charge density $Z(s)$ via the equation

$$Z(s) = \sum_j Z_j \delta(s - s_j), \tag{8}$$

so that the volume charge density $\rho(\mathbf{r},t)$ is [17]

$$\rho(\mathbf{r},t) = \iiint Z(s) \delta(\mathbf{r} - \mathbf{x} - s\mathbf{u}') C(\mathbf{x},\mathbf{u}',t) d^3\mathbf{x} d^2\mathbf{u}' ds, \tag{9}$$

where the function $C$ denotes the microscopic concentration of particles with position $\mathbf{r}$ and orientation $\mathbf{u}$ at time $t$ defined by

$$C(\mathbf{r},\mathbf{u},t) = \sum_{i=1}^{N} \delta(\mathbf{r} - \mathbf{r}_i) \delta[\mathbf{u} - \mathbf{u}_i(t)]. \tag{10}$$

Again, in close analogy with electrostatics, a particle having dipole moment orientation $\mathbf{u}$ and located at $\mathbf{r}$ experiences a magnetic field arising from the interaction potential energy

$$U_{ddi}(\mathbf{r},\mathbf{u},t) = \int Z(s) \Psi(\mathbf{r} + s\mathbf{u},t) ds \tag{11}$$

superimposed on the anisotropy energy of the particle and the energy arising from externally applied fields. Thus, the magnetization of the assembly $\mathbf{M}(t)$ may be written

$$\mathbf{M}(t) = \frac{M_s}{N} \sum_{i=1}^{N} \mathbf{u}_i(t) \tag{12}$$

where $M_s$ is the saturation magnetization of the material, and may be linked to $C$ by the equation

$$\mathbf{M}(t) = \frac{M_s}{N} \int_{V_s} \int_{\Omega_\mathbf{u}} \mathbf{u} C(\mathbf{r},\mathbf{u},t) d^3\mathbf{r} d^2\mathbf{u}, \tag{13}$$

where the integrations must be taken over $V_s$ the volume of the sample, and $\Omega_\mathbf{u}$ the total space angle swept by unit vector $\mathbf{u}$. We note that Eqs.(1)-(13) are model-independent, with the exception of the assumptions of identical volumes and frozen mechanical motions.

Thus in order to estimate the effect of dipole-dipole interactions on the thermal fluctuations of the magnetization of interacting single-domain ferromagnetic particles, one requires the time evolution for $C(\mathbf{r},\mathbf{u},t)$. This cannot be done without an equation of motion specifying the dynamics of $\mathbf{u}_i(t)$.

### III. EQUATION OF MOTION FOR THE DISTRIBUTION FUNCTION OF MAGNETIZATION ORIENTATIONS

Here we derive of an equation of motion for the distribution function of magnetization orientations, including the effect of thermal fluctuations. In order to accomplish this, one requires an equation specifying the dynamics of each $\mathbf{u}_i(t)$. Since the magnetization of a particle is subjected to thermal agitation, one may take as equations of motion the set of Gilbert-Langevin equations for $\mathbf{u}_i(t)$, viz.

$$\dot{\mathbf{u}}_i(t) = \frac{\gamma_{\textit{eff}}}{M_s}\left[\mathbf{u}_i(t) \times \mathbf{H}_i^{\textit{eff}}\right] - \frac{\alpha \gamma_{\textit{eff}}}{M_s}\mathbf{u}_i(t) \times \left[\mathbf{H}_i^{\textit{eff}} \times \mathbf{u}_i(t)\right], \tag{14}$$

where $\alpha$ is a dimensionless damping constant, $\mathbf{u}_i$ is a unit vector along the magnetization of particle $i$, $\gamma_{\textit{eff}} = \gamma/(1+\alpha^2)$, $\gamma$ is the gyromagnetic ratio, and

$$\mathbf{H}_i^{\textit{eff}} = -\frac{\partial U}{\partial \mathbf{u}_i} + \mathbf{h}_i(t), \tag{15}$$

is the effective magnetic field acting on particle *i*. In the above equation, $\mathbf{h}_i(t)$ is the Gaussian random white noise field acting on *i* and has the following properties [3]

$$\overline{\mathbf{h}_i(t)} = \mathbf{0}, \tag{16}$$

$$\overline{h_{i,\mu}(t) h_{j,\kappa}(t')} = 2\delta_{ij}\delta_{\mu\kappa} \frac{\alpha}{\gamma \beta M_s v} \delta(t-t'), \tag{17}$$

where the overbar denotes an average over the distribution of realizations of the white noises [5] while $\beta = 1/kT$, $\delta_{ij}$ is Kronecker's delta, $v$ is the volume of a particle, $T$ is the absolute temperature, and $U$ is the free energy made of local anisotropy energies, Zeeman energies and the dipole-dipole interaction between the particles.

Since the time behavior of each $\mathbf{u}_i(t)$ is governed by a Langevin equation (14), it follows that each $\mathbf{u}_i(t)$ is a time-dependent random variable. It also follows that $C(\mathbf{r},\mathbf{u},t)$, and, therefore, $\rho(\mathbf{r},t)$, $\Psi(\mathbf{r},t)$ and $U_{ddi}(\mathbf{r},\mathbf{u},t)$ are functions of time-dependent random variables so that instead of the evolution equation for $C(\mathbf{r},\mathbf{u},t)$, we seek an equation of motion for the single-particle distribution function, $W(\mathbf{r},\mathbf{u},t) = \overline{C(\mathbf{r},\mathbf{u},t)}$. Following Brown [3] and Berne [17], the equation that governs the dynamics of $W(\mathbf{r},\mathbf{u},t)$ is the Fokker-Planck equation

$$2\tau_N \frac{\partial W}{\partial t} = \frac{\beta}{\alpha} \mathbf{u} \cdot \left[ \overline{\frac{\partial U}{\partial \mathbf{u}} \times \frac{\partial C}{\partial \mathbf{u}}} \right] + \frac{\partial}{\partial \mathbf{u}} \cdot \left[ \overline{\beta C \frac{\partial U}{\partial \mathbf{u}}} + \frac{\partial W}{\partial \mathbf{u}} \right] \tag{18}$$

where $\tau_N = \beta(1+\alpha^2) M_s v / (2\alpha\gamma)$ is the free diffusion time, and

$$U(\mathbf{r},\mathbf{u},t) = U_s(\mathbf{r},\mathbf{u},t) + U_{ddi}(\mathbf{r},\mathbf{u},t), \tag{19}$$

$U_s(\mathbf{r},\mathbf{u},t)$ is a single-particle potential containing the anisotropy and Zeeman energies, and $U_{ddi}(\mathbf{r},\mathbf{u},t)$ is the dipole-dipole interaction potential experienced by a particle located at **r** with orientation **u** at time *t*. We note that in assemblies, in general the easy axes of the particles are not aligned, so that strictly speaking, the anisotropy term should depend on **r**. However, this hypothesis is valid for a fully aligned system of single-domain ferromagnetic particles (that is, with easy and hard directions parallel to each other). Here, we will neglect this dependence because our objective is the effect of dipole-dipole interaction on magnetic relaxation, and replace the true anisotropy term by an effective one. From now on, we will discard the gyromagnetic term in Eq. (18), that is, the first term on the right hand side of this equation. Although these terms are important to obtain the nontrivial damping dependence of the relaxation time and also for the calculation of the short time behavior of the magnetization and ferromagnetic resonance, they will not change our basic conclusions regarding the effect of interactions. Thus, Eq. (18) becomes

$$2\tau_N \left[ \frac{\partial}{\partial t} - L_{FP} \right] W = \frac{\partial}{\partial \mathbf{u}} \cdot \left[ \overline{\beta C \frac{\partial U_{ddi}}{\partial \mathbf{u}}} \right] \quad (20)$$

where

$$L_{FP} W = (2\tau_N)^{-1} \frac{\partial}{\partial \mathbf{u}} \cdot \left[ \beta W \frac{\partial U_s}{\partial \mathbf{u}} + \frac{\partial W}{\partial \mathbf{u}} \right] \quad (21)$$

is the interaction-free single particle Fokker-Planck operator. Clearly, one may regard the dipole-dipole interaction term as a driving of the diffusive dynamics of the magnetization of a single particle.

Now, we are interested in the average magnetization, viz.

$$\mathbf{M}(t) = \frac{M_s}{N} \lim_{\mathbf{q} \to 0} \int_{V_s} \int_{\Omega_\mathbf{u}} \mathbf{u} W(\mathbf{r}, \mathbf{u}, t) e^{i\mathbf{q} \cdot \mathbf{r}} d^3\mathbf{r} d^2\mathbf{u}, \quad (22)$$

so that all the required information is contained in the zero wave vector limit of the spatial Fourier transform of $W(\mathbf{r}, \mathbf{u}, t)$. Thus, we introduce the Fourier transforms

$$C(\mathbf{q},\mathbf{u},t) = \int C(\mathbf{r},\mathbf{u},t)e^{i\mathbf{q}\cdot\mathbf{r}}d^3\mathbf{r}, \; U_{ddi}(\mathbf{q},\mathbf{u},t) = \int U_{ddi}(\mathbf{r},\mathbf{u},t)e^{i\mathbf{q}\cdot\mathbf{r}}d^3\mathbf{r}$$

$$W(\mathbf{q},\mathbf{u},t) = \int W(\mathbf{r},\mathbf{u},t)e^{i\mathbf{q}\cdot\mathbf{r}}d^3\mathbf{r} = \int \overline{C(\mathbf{r},\mathbf{u},t)}e^{i\mathbf{q}\cdot\mathbf{r}}d^3\mathbf{r}$$

and take the Fourier transform of Eq. (20). In the zero wave vector limit, Eq. (20) becomes

$$2\tau_N \left[\frac{\partial}{\partial t} - L_{FP}\right] W(\mathbf{u},t) = \beta \frac{\partial}{\partial \mathbf{u}} \cdot \mathbf{F}(\mathbf{u},t) \tag{23}$$

where

$$\mathbf{F}(\mathbf{u},t) = \overline{\int C(\mathbf{r},\mathbf{u},t)\frac{\partial U_{ddi}}{\partial \mathbf{u}}(\mathbf{r},\mathbf{u},t)d^3\mathbf{r}} = \frac{1}{8\pi^3}\overline{\int C(\mathbf{q}_1,\mathbf{u},t)\frac{\partial U_{ddi}}{\partial \mathbf{u}}(-\mathbf{q}_1,\mathbf{u},t)d^3\mathbf{q}_1}$$
$$= \frac{\mathbf{T}}{8\pi^3} \tag{24}$$

$$C(\mathbf{q},\mathbf{u},t) = \sum_{j=1}^{N} e^{i\mathbf{q}\cdot\mathbf{r}_j}\delta[\mathbf{u}-\mathbf{u}_j(t)], \; W(\mathbf{u},t) = \overline{\delta[\mathbf{u}-\mathbf{u}_i(t)]} \tag{25}$$

and $U_{ddi}(\mathbf{q},\mathbf{u},t)$ may be computed from the Fourier transform of Eqs. (6) and (11). If we assume

$$\int ds Z(s) = 0, \; \int s Z(s) ds = M_s v \text{ and } \int s^L Z(s) ds = 0, \; L \geq 2, \tag{26}$$

the interaction terms consist of nine terms rather than an infinite number (see Appendix). In these terms we have lattice sums $S_{2,i}$ defined by

$$S_{2,0} = \sum_i \sum_{j \neq i} \frac{P_2(\cos\vartheta_{ij})}{r_{ij}^3}, \; S_{2,\pm 1} = \sum_i \sum_{j \neq i} \frac{P_2^{\pm 1}(\cos\vartheta_{ij})e^{\pm i\varphi_{ij}}}{r_{ij}^3},$$

$$S_{2,\pm 2} = \sum_i \sum_{j \neq i} \frac{P_2^{\pm 2}(\cos\vartheta_{ij})e^{\pm 2i\varphi_{ij}}}{r_{ij}^3},$$

where $(\vartheta_{ij}, \varphi_{ij})$ denote the spherical polar angles specifying the orientation of vector $\mathbf{r}_{ij}$, and $P_l^m(z)$ is an associated Legendre function. Because the particles *are distributed at random* in the sample, we may replace these sums by *their spatial mean value*. These averages may be calculated by using a continuum approximation as follows. On defining the dipole-dipole interaction tensor **D** as

$$\mathbf{D}(\mathbf{r}-\mathbf{r}') = \frac{3(\mathbf{r}-\mathbf{r}')(\mathbf{r}-\mathbf{r}') - \mathbf{I}|\mathbf{r}-\mathbf{r}'|^2}{|\mathbf{r}-\mathbf{r}'|^5},$$

where **I** is the identity tensor, we have, *for magnetic dipoles*

$$\int \mathbf{D}(\mathbf{r}-\mathbf{r}') d^3\mathbf{r}' = \frac{8\pi}{3}\mathbf{I}, \tag{27}$$

since the integral of the dipolar magnetic *induction* over an infinite volume is $8\pi\mathbf{m}/3$, where **m** is the magnetic dipole moment of the volume and where we have used Gaussian CGS units [34]. From Eq. (27) it immediately follows that $S_{2,\pm 1}$ and $S_{2,\pm 2}$ are zero since they correspond to off-diagonal elements of **D**, while $S_{2,0} = 4\pi\rho_0/3$, where $\rho_0 = N/V_s$ is the mean number of particles per unit volume of the sample. In the uniaxial approximation, we may write Eq. (23) as follows

$$2\tau_N \sin\vartheta \left[\frac{\partial}{\partial t} - L_{FP}\right] W = \beta \frac{\partial}{\partial \vartheta}\left[\sin\vartheta W \frac{\partial V_{ddi}}{\partial \vartheta}\right], \tag{28}$$

where $(\vartheta, \varphi)$ is the set of spherical polar angles specifying the orientations of **u**, $\beta V_{ddi}$ is

$$\beta V_{ddi}(\vartheta, t) = -\lambda \cos\vartheta \langle \cos\vartheta \rangle(t). \tag{29}$$

and $\lambda$ is the interaction parameter given by

$$\lambda = \frac{8\pi\beta\rho_0 M_s^2 v^2}{3}, \tag{30}$$

Since we have neglected the gyroscopic terms in the Fokker-Planck equation (18) giving rise to ferromagnetic resonance, the nonaxially-symmetric character of the single particle free energy $U_s(\mathbf{u}, t)$ becomes irrelevant in order to give a first (rough) estimate of the effect of dipole-dipole interactions on quantities pertaining to longitudinal magnetic relaxation. Thus, for simplicity, we will take $U_s(\mathbf{u},t) = U_s(\vartheta,t)$ so that Eq. (28) radically simplifies into the axially symmetric equation for the probability density function $W(\vartheta,t)$

$$2\tau_N \sin\vartheta \frac{\partial}{\partial t} W = \frac{\partial}{\partial \vartheta}\left[\sin\vartheta\left(\beta W \frac{\partial V}{\partial \vartheta} + \frac{\partial W}{\partial \vartheta}\right)\right], \qquad (31)$$

where

$$V(\vartheta,t) = U_s(\vartheta,t) + V_{ddi}(\vartheta,t).$$

In the following, we will focus on this axially symmetric equation in order to obtain the equilibrium magnetization and relaxation time for an assembly of uniaxial particles.

## IV. CALCULATION OF THE EQUILIBRIUM MAGNETIZATION

In this paragraph and later in the text, we take $U_s(\vartheta,t)$ in Eq. (31) as a uniaxial anisotropy energy, and assume that the easy axis, the dipole field and external field are all aligned. Thus, Eq. (31) becomes

$$2\tau_N \frac{\partial W}{\partial t} = \frac{\partial}{\partial z}\left\{(1-z^2)\left[\beta W \frac{\partial U}{\partial z} + \frac{\partial W}{\partial z}\right]\right\}, \qquad (32)$$

where $z = \cos\vartheta$ and

$$\beta U(z,t) = \beta U_s(z) - \lambda z \int_{-1}^{1} z' W(z',t) dz', \qquad (33)$$

exemplifying the nonlinear nature of Eq. (32). Furthermore, we take

$$\beta U_s(z) = -\sigma z^2 - \xi z, \qquad (34)$$

where $\sigma = \beta K_{eff} v$, $\xi = \beta M_s H v$, $K_{eff}$ is an effective anisotropy constant, and $H$ is the intensity of the externally applied field. In equilibrium, $\partial_t W = 0$ and therefore, the equilibrium solution $W_0(z)$ of Eq. (32) obeys the nonlinear integral equation

$$W_0(z) = Z^{-1} e^{-\beta U_0(z)}, \tag{35}$$

where

$$\begin{aligned}\beta U_0(z) &= \beta U_s(z) - \lambda z \int_{-1}^{1} z' W_0(z') dz' \\ &= \beta U_s(z) - \lambda z \langle z \rangle_0,\end{aligned} \tag{36}$$

$$Z = \int_{-1}^{1} e^{-\beta U_0(z)} dz.$$

Because we are interested in the equilibrium magnetization only, the only required quantity is $\langle z \rangle_0$, to which the equilibrium magnetization is proportional (all mean transverse components are zero in this case). If $\sigma \to 0$ (high temperature limit), $\langle z \rangle_0$ obeys the self-consistent equation

$$\langle z \rangle_0 = \coth\left[\xi + \lambda \langle z \rangle_0\right] - \left(\xi + \lambda \langle z \rangle_0\right)^{-1}, \tag{37}$$

which must be solved numerically for arbitrary $\xi$. Indeed, the only physically acceptable value for $\langle z \rangle_0$ as $\xi \to 0$ is $\langle z \rangle_0 = 0$ because the dipolar field is insufficient to orient the whole assembly. Thus for $\xi \ll 1$, we also have $\left(\xi + \lambda \langle z \rangle_0\right) \ll 1$ so that replacing the right hand side of Eq. (37) by its Taylor series expansion, we have

$$\langle z \rangle_0 \approx \frac{\xi}{3 - \lambda}, \quad \xi \ll 1. \tag{38}$$

This clearly demonstrates that as $\sigma \to 0$, the limiting value of $\lambda$ to be taken is $\lambda \leq 3$.

As $\sigma \to \infty$ (high energy barrier), Eq. (37) must be replaced by

$$\langle z \rangle_0 = \operatorname{th}\left[\xi + \lambda \langle z \rangle_0\right], \tag{39}$$

which yields, in the low field limit

$$\langle z \rangle_0 \approx \frac{\xi}{1-\lambda}, \quad \xi \ll 1, \tag{40}$$

indicating that the range of validity of the model is $\lambda \leq 1$ at large barriers. In cases intermediate between no anisotropy and very large anisotropy, $\langle z \rangle_0$ must be calculated numerically from the equation

$$\langle z \rangle_0 = \frac{1}{Z} \int_{-1}^{1} z W_0(z) dz. \tag{41}$$

This may be accomplished by using some numerical root-finding technique. Now we calculate the superparamagnetic relaxation time of the assembly.

## V. CALCULATION OF THE SUPERPARAMAGNETIC RELAXATION TIME

Here, we focus on the numerical calculation of the thermally activated time scale of the assembly from the time-dependent solution of Eq. (32). Because we assume that the interaction term is small, we may seek the solution of Eq. (32) via perturbation theory :

$$W(z,t) = W_0(z) + w(z,t), \quad w(z,t) << W_0(z), \tag{42}$$

so that Eq. (32) becomes

$$2\tau_N \frac{\partial w}{\partial t} = \frac{\partial}{\partial z}\left[(1-z^2)\left(\frac{\partial w}{\partial z} + \beta w \frac{\partial U_0}{\partial z} + \beta W_0 \frac{\partial U_1}{\partial z}\right)\right], \tag{43}$$

where

$$\beta U_1(z,t) = -\lambda z \int_{-1}^{1} z' w(z',t) dz'. \tag{44}$$

We then expand both $W_0(z)$ and $w(z,t)$ in a series of Legendre polynomials, viz.

$$W_0(z) = \sum_{n=0}^{\infty} (n+1/2) f_n P_n(z), \tag{45}$$

$$w(z,t) = \sum_{m=1}^{\infty}(m+1/2)g_m(t)P_m(z). \qquad (46)$$

From the orthogonality of the $P_n$, we have

$$f_n = \langle P_n \rangle_0$$

where the angular brackets $\langle \ \rangle_0$ denotes an average over $W_0(z)$. Then we combine Eqs. (45) and (46) with Eq.(43), and use the recurrence and orthogonality properties of the Legendre polynomials in order to obtain the hierarchy of differential recurrence relations

$$\frac{2\tau_N}{n(n+1)}\dot{g}_n(t) + \left[1 - \frac{2\sigma}{(2n-1)(2n+3)}\right]g_n(t) = \frac{(\xi + \lambda f_1)}{(2n+1)}\left[g_{n-1}(t) - g_{n+1}(t)\right]$$
$$+2\sigma\left[\frac{n-1}{(2n-1)(2n+1)}g_{n-2}(t) - \frac{n+2}{(2n+1)(2n+3)}g_{n+2}(t)\right] + \frac{\lambda g_1(t)}{(2n+1)}[f_{n-1} - f_{n+1}].$$

(47)

Notice that the (linearized) Warchol-Vaughan hierarchy may be obtained from the above one by setting $\sigma = \xi = 0$ and formally replacing $\lambda$ by $-\lambda$.

In order to compute the relaxation time, one may arrange the set of Eqs. (47) in matrix form, namely

$$\dot{\mathbf{X}} = \mathbf{A}\mathbf{X} \qquad (48)$$

where $\mathbf{X}$ is the column vector

$$\mathbf{X} = \begin{pmatrix} g_1(t) \\ g_2(t) \\ g_3(t) \\ \vdots \\ \vdots \\ \vdots \\ \vdots \end{pmatrix}$$

and $\mathbf{A}$ is given by

$$\mathbf{A} = -\frac{1}{\tau_N} \begin{pmatrix} 1-\frac{2\sigma}{5}+\frac{\lambda}{3}[1-f_2] & \frac{\xi+\lambda f_1}{3} & \frac{2\sigma}{5} & 0 & 0 & \cdots \\ -\frac{3(\xi+\lambda f_1)}{5}+\frac{3\lambda}{5}(f_1-f_3) & 3-\frac{2\sigma}{7} & \frac{3(\xi+\lambda f_1)}{5} & \frac{24\sigma}{35} & 0 & \cdots \\ -\frac{24\sigma}{35}+\frac{6\lambda}{7}(f_2-f_4) & -\frac{6(\xi+\lambda f_1)}{7} & 6-\frac{4\sigma}{15} & \frac{6(\xi+\lambda f_1)}{7} & \frac{20\sigma}{21} & 0 & \cdots \\ \frac{10\lambda}{9}(f_3-f_5) & -\frac{20\sigma}{21} & -\frac{10(\xi+\lambda f_1)}{9} & 10-\frac{20\sigma}{77} & \frac{10(\xi+\lambda f_1)}{9} & \frac{40\sigma}{33} & 0 \\ \vdots & \vdots & \vdots & \vdots & \vdots & \vdots & \vdots \\ \vdots & \vdots & \vdots & \vdots & \vdots & \vdots & \vdots \end{pmatrix}.$$

Clearly, the matrix **A** depends on *equilibrium values* of *all* the Legendre polynomials. The matrix **A** may be diagonalized numerically to compute its smallest nonvanishing eigenvalue $\Lambda_1$, the inverse of which yields the thermally activated relaxation time of the magnetization of the assembly $\tau$. Now, we may also derive an expression for the thermally activated relaxation rate in the high energy barrier limit by using a variational procedure which we describe in the following section.

## VI. ASYMPTOTIC EXPRESSIONS FOR THE RELAXATION RATE

In order to derive an asymptotic formula for $\Lambda_1$ valid in the high anisotropy barrier limit, we substitute

$$w(z,t) = \phi(z,t)e^{-\beta U_0(z)} + \lambda z Z^{-1} e^{-\beta U_0(z)} \int_{-1}^{1} z'\phi(z',t)e^{-\beta U_0(z')}dz'. \tag{49}$$

into Eq. (43). This equation replaces the detailed balance condition used by Brown [3] and Kramers [35] in the absence of interactions. On combining Eqs. (43) and (49), we obtain

$$2\tau_N e^{-\beta U_0(z)}\left[\frac{\partial \phi}{\partial t} - \beta Z^{-1}\frac{\partial U_1}{\partial t}\right] = \frac{\partial}{\partial z}\left[(1-z^2)e^{-\beta U_0(z)}\frac{\partial \phi}{\partial z}\right]. \tag{50}$$

On setting $\phi(z,t) = \psi(z)e^{-\Gamma t}$ in Eq. (50), multiplying the resulting equation by $\psi(z)$ and integrating both sides between -1 and 1, we obtain the variational equation for the relaxation rate $\Gamma$, viz.

$$2\Gamma\tau_N = \frac{\int_{-1}^{1}(1-z^2)e^{-\beta U_0(z)}\psi'^2(z)dz}{\int_{-1}^{1}\psi^2(z)e^{-\beta U_0(z)}dz + \lambda Z^{-1}\int_{-1}^{1}\int_{-1}^{1}z\psi(z)e^{-\beta U_0(z)}z'\psi(z')e^{-\beta U_0(z')}dzdz'} \quad (51)$$

with the constraints

$$\int_{-1}^{1}\psi(z)e^{-\beta U_0(z)}dz = -\lambda Z^{-1}\int_{-1}^{1}\int_{-1}^{1}ze^{-\beta U_0(z)}z'\psi(z')e^{-\beta U_0(z')}dzdz'$$

$$\int_{-1}^{1}\psi^2(z)e^{-\beta U_0(z)}dz + \lambda Z^{-1}\int_{-1}^{1}\int_{-1}^{1}z\psi(z)e^{-\beta U_0(z)}z'\psi(z')e^{-\beta U_0(z')}dzdz' = const \quad (52)$$

showing that the eigenfunctions of the linearized Fokker-Planck equation (43) are, in general, *not orthogonal* to the equilibrium distribution $W_0$. Because the denominator in Eq. (51) is a constant, the right hand side of Eq. (51) is stationary if $\psi$ obeys the Euler-Lagrange equation

$$\frac{d}{dz}\left[(1-z^2)e^{-\beta U_0}\frac{d\psi}{dz}\right] = 0$$

which is exactly the same equation as Brown's Eq. (4.50) in Ref. 3. We may then apply all Brown's estimates in the calculation of the integrals in Eq. (51). For $\xi = 0$, this equation yields, in the high energy barrier limit

$$\Lambda_1\tau_N \approx \frac{2\sigma^{3/2}[1-\lambda]}{\sqrt{\pi}}e^{-\sigma}, \quad (53)$$

which is Brown's equation [3] multiplied by the factor $(1-\lambda)$. Moreover, for $0 < \lambda < 1$, the relaxation rate *decreases*, leading to an *increase* in the blocking temperature of the sample.

This formula renders a very large relaxation time value as $\lambda$ approaches unity, illustrating again that the proposed model fails at and above this limiting value of $\lambda$.

For $\xi \neq 0$, the relaxation rate depends on the equilibrium value $\langle z \rangle_0$, and Eq. (51) yields

$$\Gamma \tau_N = \kappa \frac{\sigma^{3/2}}{\sqrt{\pi}} (1-h^2) \left[ (1-\lambda \langle z \rangle_0)(1+h) e^{-\sigma(1+h)^2} + (1+\lambda \langle z \rangle_0)(1-h) e^{-\sigma(1-h)^2} \right], \quad (54)$$

where

$$h = \frac{\xi + \lambda \langle z \rangle_0}{2\sigma},$$

$$\kappa \approx \frac{(1+\lambda \langle z \rangle_0)(1-h) e^{2\sigma h} + (1-\lambda \langle z \rangle_0)(1+h) e^{-2\sigma h}}{M_+ + M_-} \left[ 1 - \frac{H}{M_+ + M_-} \right],$$

and

$$M_\pm = (1 \pm \lambda \langle z \rangle_0)^2 (1 \mp h) \left[ e^{\pm 2\sigma h} \mp \frac{\lambda(1 \mp h)}{D} \right],$$

$$H = \frac{4\lambda (1 - \lambda^2 \langle z \rangle_0^2)}{D},$$

$$D = (1-h) e^{2\sigma h} + (1+h) e^{-2\sigma h}.$$

For $\lambda = 0$, Eq. (54) reduces to Brown's result [3] for the relaxation rate, namely

$$\Gamma \tau_N = \frac{\sigma^{3/2}}{\sqrt{\pi}} (1-h^2) \left[ (1+h) e^{-\sigma(1+h)^2} + (1-h) e^{-\sigma(1-h)^2} \right],$$

and also reduces to Eq. (53) for $\xi = 0$ because in zero field, $\langle z \rangle_0 = 0$ for $\lambda < 1$.

## VII. RESULTS AND DISCUSSION

First, we notice an apparent difference between the recurrence relations derived by Warchol and Vaughan [19] for electric dipoles and ours, Eq. (47) concerning the sign of the interaction term. However, this may be explained by the difference in the physical nature of the electric and magnetic dipole-dipole interactions. For magnetic dipoles the integral of the interaction tensor is provided by Eq. (27), while for electric ones the corresponding equation reads

$$\int \mathbf{D}(\mathbf{r}-\mathbf{r}')d^3\mathbf{r}' = -\frac{4\pi}{3}\mathbf{I}.$$

so that $S_{2,0} = -2\pi\rho_0/3$ is now negative, and for identical electric dipoles of magnitude $m$, Eq. (29) becomes $\lambda = -4\pi\beta\rho_0 m^2/3$. Thus for electric dipole-dipole interactions, we have

$$\beta V_{ddi}(\vartheta,t) = |\lambda|\cos\vartheta\langle\cos\vartheta\rangle(t),$$

in agreement with Warchol and Vaughan [19], allowing one in particular to reproduce Berne's result for the dipole autocorrelation function pertaining to dielectric relaxation of dipolar molecules.

Now, from Eq. (41), we may numerically calculate the equilibrium magnetization as a function of the applied field. The behavior of the equilibrium magnetization as a function of the applied field is shown for various values of the interaction parameter $\lambda$ on Figures 1. Clearly, the normalized equilibrium magnetization is substantially affected by the finite dilution effect, and is larger than for infinite dilution. This holds true for any value of the anisotropy. Here, however, a word of caution is necessary. We note, having used Eq. (27) to evaluate the lattice sums, that the complete dipole-dipole interaction potential should read

$$\beta V_{ddi}(\vartheta,\varphi,t) = -\lambda\cos\vartheta\langle\cos\vartheta\rangle(t)$$
$$+\frac{\lambda}{2}\left[\sin\vartheta\cos\varphi\langle\sin\vartheta\cos\varphi\rangle(t) + \sin\vartheta\sin\varphi\langle\sin\vartheta\sin\varphi\rangle(t)\right]. \quad (55)$$

so that the effect of a field applied in the direction perpendicular to the (effective) anisotropy will cause the equilibrium magnetization to be reduced in comparison to the non-interacting case, so exhibiting a behavior differing from the one shown in Figs. 1. Nevertheless, for the simple uniaxial free energy given by Eq. (34), $\langle\sin\vartheta\cos\varphi\rangle_0$ and $\langle\sin\vartheta\sin\varphi\rangle_0$ vanish irrespective of the field applied along the *effective* anisotropy axis. Therefore, the resulting effect of the dipole-dipole interactions on the magnetization curves is to enhance the equilibrium magnetic response at least in the very simple geometry of the free energy considered here, this being due to a much stronger contribution of the longitudinal component of the dipolar field than the transverse ones. However it is possible that using a uniform distribution of easy axes our calculation will render a magnetization curve similar to the result of Chantrell *et al*. [36].

Information on the nonequilibrium properties of the assembly of single-domain ferromagnetic particles is mostly contained in the inverse of the smallest nonvanishing eigenvalue of the matrix **A** in Eq. (48) (longest relaxation time). One may *a priori* expect some dependence of this time scale on the equilibrium values of all Legendre polynomials. An interesting feature of our Eq. (53) is that when the two wells of the effective free energy are equivalent, the relaxation time *does not depend on these values at all*. As Figure 2 demonstrates, it is indeed clear that at high energy barriers (meaning $\sigma \geq 2$), this is true, since comparison between the numerical and asymptotic calculations exhibits perfect agreement. As the dc field is increased, the longest relaxation time depends only on the equilibrium magnetization, and not on the higher order Legendre polynomials. This is again demonstrated in Figure 3. Hence for ZFC experiments in moderate fields, the equilibrium state becomes an important feature in the interpretation of experimental data, since the peak of the ZFC temperature maximum as a function of the applied field disappears as the concentration increases.

Next, it is worth comparing our result Eq. (53) with the calculation rendered by Shiino's perturbation theory [37]. Indeed, our Eq. (53) is similar to Shiino's Eq. (5.27), but obtained by another method. Our Eq. (54) extends Shiino's result to asymmetric bistable potentials, which he did not consider in his paper. Actually, Shiino's equation as applied to our problem renders the integral relaxation time which has been shown to diverge exponentially from the thermally activated relaxation time as the asymmetry of the wells is increased. This effect was discovered in the context of magnetic relaxation of *noninteracting* single-domain ferromagnetic particles by Coffey *et al.* [38], and was later explained by Garanin [39] as a consequence of the depletion of the shallower well by the biasing effect of the symmetric bistable potential. Clearly, this effect should also arise in our model.

Thus, in this paper, we have proposed a model allowing one to handle thermal relaxation of an assembly of single-domain ferromagnetic particles in the presence of weak dipole-dipole interactions. To accomplish this, we have solved the nonlinear FPE (28) and obtained the thermally activated relaxation time of the assembly, Eqs. (53) and (54). In particular, our Eq. (54) constitutes a new result for the relaxation time of assemblies in a uniform external field. Furthermore, the model may be in principle extended in several ways, provided that the concentration of magnetic matter is not too large. Of course, the extension of the proposed model to higher concentrations requires the inclusion of orientational correlations in the calculation. The treatment then becomes mathematically much more involved.

## APPENDIX : HANDLING THE INTERACTION TERM

Here we give some steps in handling the interaction term in Eq. (23). First, the Fourier transform of the interaction potential is

$$U_{ddi}(\mathbf{q},\mathbf{u},t) = \frac{4\pi}{q^2} \iiint Z(s)Z(s')C(\mathbf{q},\mathbf{u}',t)e^{i\mathbf{q}\cdot(s'\mathbf{u}'-s\mathbf{u})}d^2\mathbf{u}'ds'ds. \qquad (A.1)$$

Now we need to explicit **T** in Eq. (24). We have

$$\mathbf{T} = 4\pi \sum_{j=1}^{N}\sum_{k=1}^{N} \iiiint e^{-i\mathbf{q}_1\cdot s'\mathbf{u}'}\partial_{\mathbf{u}}\left[e^{i\mathbf{q}_1\cdot s\mathbf{u}}\right]Z(s)Z(s')e^{i\mathbf{q}_1\cdot \mathbf{r}_{jk}}W_2(\mathbf{u},\mathbf{u}',t)\frac{d^3\mathbf{q}_1}{q_1^2}d^2\mathbf{u}'ds'ds$$
$$= \mathbf{T}[W_2] \qquad (A.2)$$

where $W_2(\mathbf{u},\mathbf{u}',t) = \overline{\delta[\mathbf{u}-\mathbf{u}_j(t)]\delta[\mathbf{u}'-\mathbf{u}_k(t)]}$ is the orientational pair correlation function and $\mathbf{r}_{jk} = \mathbf{r}_j - \mathbf{r}_k$. This equation may now be split in two parts, namely

$$\mathbf{T}[W_2] = 4\pi(\mathbf{T}_1 + \mathbf{T}_2)[W_2]$$

with

$$\mathbf{T}_1[W_2] = \sum_{j=1}^{N} \int \frac{d^3\mathbf{q}_1}{q_1^2} \int ds Z(s) \partial_{\mathbf{u}}\left[e^{i\mathbf{q}_1 \cdot s\mathbf{u}}\right] \int ds' Z(s') \int d^2\mathbf{u}' W_2(\mathbf{u},\mathbf{u}',t) e^{-i\mathbf{q}_1 \cdot s'\mathbf{u}'} \quad (A.3)$$

representing the contribution of the contact term, and

$$\mathbf{T}_2[W_2] = \sum_{j=1}^{N} \sum_{\substack{k=1 \\ k \neq j}}^{N} \int \frac{d^3\mathbf{q}_1}{q_1^2} e^{i\mathbf{q}_1 \cdot \mathbf{r}_{jk}} \int ds Z(s) \partial_{\mathbf{u}}\left[e^{i\mathbf{q}_1 \cdot s\mathbf{u}}\right] \int ds' Z(s') \int d^2\mathbf{u}' W_2(\mathbf{u},\mathbf{u}',t) e^{-i\mathbf{q}_1 \cdot s'\mathbf{u}'} \quad (A.4)$$

representing the contribution of the pair interaction term. We shall show that $\mathbf{T}_1[W_2]$ does not contribute at all. To this purpose, we use the Rayleigh expansion

$$e^{i\mathbf{q}\cdot\mathbf{r}} = 4\pi \sum_{J=0}^{\infty} \sum_{K=-J}^{J} i^J j_J(qr) Y_{JK}(\hat{r}) Y_{JK}^*(\hat{q}) \quad (A.5)$$

and replace each exponential in Eq. (A.3) so that we obtain

$$\mathbf{T}_1[W_2] = (4\pi)^2 N \sum_{JKLM} i^{J-L} \partial_{\mathbf{u}} Y_{JK}(\mathbf{u}) \times$$

$$\int d^2\mathbf{u}' W_2(\mathbf{u},\mathbf{u}',t) Y_{LM}^*(\mathbf{u}') \int ds Z(s) \int ds' Z(s') \int_0^{\infty} j_J(q_1 s) j_L(q_1 s') dq_1 \delta_{JL} \delta_{KM} \quad (A.6)$$

where $d^3\mathbf{q}_1 = q_1^2 dq_1 d^2\hat{q}_1$ and the orthogonality property of the spherical harmonics in $\mathbf{q}$ space have been used. Now, we have [40]

$$\int_0^\infty j_J(q_1 s) j_J(q_1 s') dq_1 = \frac{1}{2(2J+1)} \begin{cases} \dfrac{\pi s^J}{s'^{J+1}}, & s < s' \\ \dfrac{\pi s'^J}{s^{J+1}}, & s' < s \\ \dfrac{\pi}{s}, & s = s' \end{cases} \qquad (A.7)$$

so that in Eq. (A.6) moments of the linear magnetic pole density like $\int s^{-J} Z(s) ds$ with positive $J$ will occur. Now, the linear magnetic pole density cannot have such moments as they have no meaning (this actually means that the magnetic volume charge distribution has no moment of negative order). Also, since magnetic monopoles do not exist, we also have $\int Z(s) ds = 0$. Therefore, $\mathbf{T}_1[W_2] = \mathbf{0}$ and $\mathbf{T}[W_2] = 4\pi \mathbf{T}_2[W_2]$.

Thus it is clear that Eq. (23) is not a closed equation for $W(\mathbf{u},t)$. The solution of this equation requires a knowledge of the orientational pair correlation function $W_2(\mathbf{u},\mathbf{u}',t)$, therefore one must find an equation that determines $W_2(\mathbf{u},\mathbf{u}',t)$. Throughout this work we choose the *simplest* possible assumption known in the field of nonequilibrium statistical mechanics for the orientational pair correlation function consisting of the neglect of orientational correlations between pair of particles. Thus, we assume that the orientational pair correlation function factorizes (mean field-like approximation) so that

$$W_2(\mathbf{u},\mathbf{u}',t) \approx W(\mathbf{u},t) W(\mathbf{u}',t) \qquad (A.8)$$

Thus, this approximation allows Eq. (20) to be closed.

We now may continue by making $\mathbf{T}_2$ more explicit. This may be accomplished again using Eq. (A.5) in conjunction with Eq. (A.4). After tedious algebra, we obtain

$$\mathbf{T}_2 = (4\pi)^3 W(\mathbf{u},t) \sum_{\substack{j,k \\ k \neq j}} \sum_{\substack{JK \\ LM \\ PQ}} i^{J-L+P} I_{JLP}(r_{jk}) K_{JKLMPQ} [\partial_{\mathbf{u}} Y_{JK}(\mathbf{u})] Y_{PQ}(\hat{r}_{jk}) \times$$
$$\int d^2\mathbf{u}' W(\mathbf{u}',t) Y_{LM}^*(\mathbf{u}'),$$
(A.9)

where

$$I_{JLP}(r_{jk}) = \int ds Z(s) \int ds' Z(s') \int_0^\infty j_P(q_1 r_{jk}) j_J(q_1 s) j_L(q_1 s') dq_1,$$
(A.10)

$$K_{JKLMPQ} = \int Y_{PQ}^*(\hat{q}_1) Y_{JK}^*(\hat{q}_1) Y_{LM}(\hat{q}_1) d^2\hat{q}_1.$$
(A.11)

These two integrals may again be evaluated with the help of Tables [40]. We have

$$I_{JLP}(r_{jk}) = \frac{\pi^{3/2}}{8 r_{jk}^{L+J+1}} \frac{\Gamma[(L+J+P+1)/2]}{\Gamma(J+3/2)\Gamma(L+3/2)\Gamma[1-(L+J-P)/2]}$$
$$\times \int ds Z(s) s^J \int ds' Z(s') s'^L F_4\left[\frac{J+L-P}{2}, \frac{J+L+P+1}{2}; J+\frac{3}{2}, L+\frac{3}{2}; \frac{s^2}{r_{jk}^2}, \frac{s'^2}{r_{jk}^2}\right],$$
(A.12)

where $\Gamma$ is the Euler gamma function [41], $F_4$ is the two-variable hypergeometric series [40]

$$F_4(\gamma, \delta; \varepsilon, \zeta; x, y) = \sum_{m=0}^\infty \sum_{n=0}^\infty \frac{(\gamma)_{m+n} (\delta)_{m+n}}{(\varepsilon)_m (\zeta)_m m! n!} x^m y^n,$$

$$K_{JKLMPQ} = \sqrt{\frac{(2J+1)(2P+1)}{4\pi(2L+1)}} \langle P0J0|L0\rangle \langle PQJK|LM\rangle,$$

the Pochhammer symbol is given by $(a)_n = \Gamma(a+n)/\Gamma(a)$, and $\langle PQJK|LM\rangle$ is a Clebsch-Gordan coefficient. We note in passing that the hypergeometric series in Eq. (A.12) converges only if [40] $s + s' < r_{jk}$, which means that the theory has a meaning only if the mean interparticle distance is larger than the particle diameter.

**ACKNOWLEDGEMENTS**

The author is grateful to Yu. P. Kalmykov and W. T. Coffey for their reading of the manuscript. Financial support of the ECO-NET program (project n°21394NH) is also gratefully acknowledged.

**FIGURE CAPTIONS**

Figure 1 . Equilibrium magnetization curves for various values of the parameter $\lambda$ calculated from Eq. (41).

Figure 2 . The relaxation time vs $\sigma$ as obtained from the diagonalization of the matrix **A** in Eq. (48) (solid line) and calculated with Eq. (53) (dots) for various values of the parameter $k_0 = \lambda/\sigma$ for $\xi = 0$.

Figure 3 . The relaxation time vs $\sigma$ as obtained from the diagonalization of the matrix **A** in Eq. (48) (solid line) and calculated with Eq. (54) (dots) for $k_0 = 0.05$ and various values of $h_0 = \xi/(2\sigma)$. The dots on curve 1 are computed with Eq. (53).

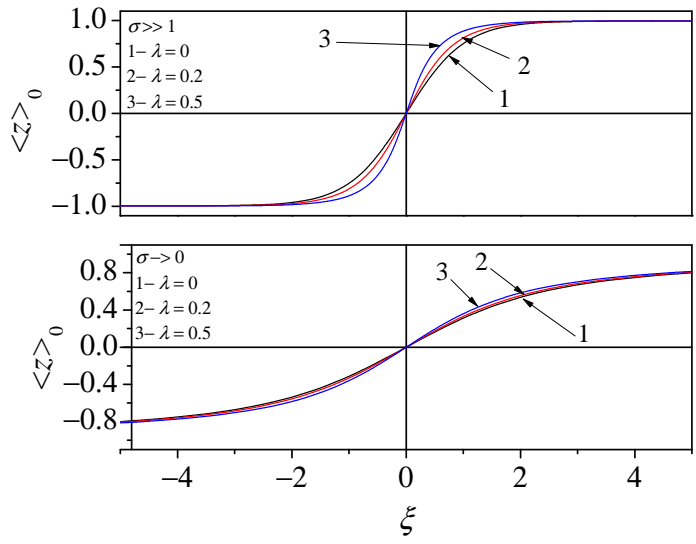

Figure 1

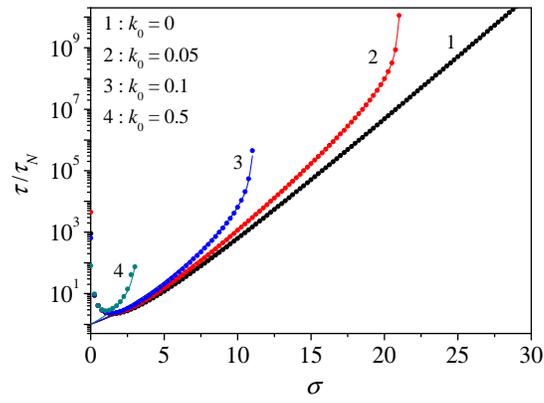

Figure 2

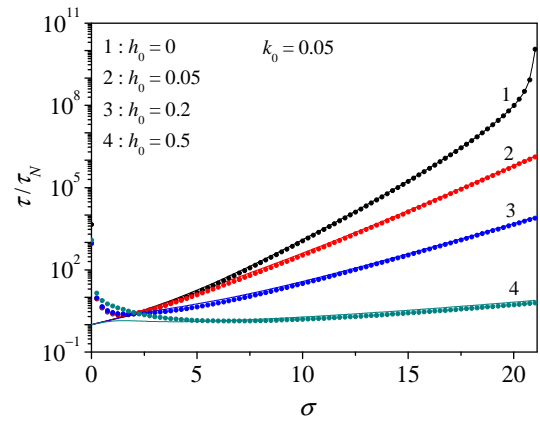

Figure 3